\documentclass[amsmath,trackchanges,twocolumn]{aastex702}

\usepackage{graphicx}
\usepackage{epsfig}
\usepackage{ulem}
\usepackage{morefloats}
\usepackage{rotating}
\usepackage{xcolor,ulem}
\usepackage{slashed}
\usepackage{aas_macros}
\usepackage{enumitem}

\received{August 31, 2026}
\revised{X}
\accepted{Y}

\begin{document}

\title{Early galactic heavy element production - a phenomenological approach} 

\author[orcid=0009-0002-5414-9292]{B. Wehmeyer}
\affiliation{Institute of Theoretical Physics, University of Wroclaw, pl. Maxa Borna 9, 50-204 Wroclaw, Poland}
\affiliation{Centre for Astrophysics Research, University of Hertfordshire, College Lane, Hatfield AL10 9AB, UK}
 \email{benjamin.wehmeyer@uwr.edu.pl}
 
\author[orcid=0000-0002-8399-5183]{D. Blaschke}
\affiliation{Institute of Theoretical Physics, University of Wroclaw, pl. Maxa Borna 9, 50-204 Wroclaw, Poland}
\affiliation{Helmholtz-Zentrum Dresden-Rossendorf (HZDR), Bautzner Landstrasse 400, 01328 Dresden, Germany}
\affiliation{Center for Advanced Systems Understanding (CASUS), Untermarkt 20, 02826 G\"orlitz, Germany}
\email[show]{david.blaschke@uwr.edu.pl}
 
\author[orcid=0000-0002-4460-0097]{F. K. R\"{o}pke}
\affiliation{Institut f\"ur Theoretische Astrophysik, Zentrum f\"ur Astronomie der Universit\"at Heidelberg,\\ Philosophenweg 12, 69120 Heidelberg, Germany}
\affiliation{Zentrum f{\"u}r Astronomie der Universit{\"a}t Heidelberg, Astronomisches Rechen-Institut, \\
M{\"o}nchhofstr.\ 12--14, 69120 Heidelberg, Germany}
\affiliation{Heidelberger Institut f{\"u}r Theoretische Studien, Schloss-Wolfsbrunnenweg 35, 69118 Heidelberg, Germany
}
\email{friedrich.roepke@h-its.org}
 
\author[orcid=0000-0001-9319-5959]{G. R\"{o}pke}
\email{gerd.roepke@uni-rostock.de}
\affiliation{Institute of Physics, University of Rostock, Einstein-Str. 23, 18057 Rostock, Germany}

\date{\today}

\begin{abstract} 
The high abundances of heavy elements in early, extremely metal-poor stars is still an unsolved issue in Galactic Chemical Evolution (GCE). Standard approaches, such as mergers of neutron stars, have several difficulties, and several non-merger production channels are under discussion. Here, we suggest that pre-enriched inflow might explain the heavy elements seen in low metallicity stars in the Galaxy. 
To characterize the conditions of the origin of such a pre-enriched infall, we apply a heavy-element freeze-out approach.
We characterize the initial distribution by Lagrange parameters which are nonequilibrium generalizations of temperature and chemical potentials.
We then speculate about the origin of such a pre-enriched infall:  The required nonequilibrium conditions might be found in cosmological fluctuations in the early Universe, and might be connected to the formation of primordial black holes.

\end{abstract}

\section{Introduction}
Increasing multi-messenger observations gave new inputs not only to the abundances of elements in various astrophysical objects,
but also the galactic chemical evolution. 
The distribution of the light elements (up to the iron region) is  changed by (mostly hydrostatic) burning processes,
whereas about $50\%$ of the heavier elements are formed by the rapid neutron capture ($r$-process) which occurs in neutron-rich environments \citep{Burbidge:1957vc}.
For a detailed discussion of the $r$-process, we refer to the reviews of \cite{1991PhR...208..267C,2007PhR...450...97A,Cowan91,Arnould07,Thielemann11,Thielemann17,Cowan19}.
However, the dominant astrophysical sites where the observed heavy elements are created remains still under discussion {\citep[e.g.,][]{Shen15,Hirai15,Ishimaru15,Cescutti15,vandeVoort15,Wehmeyer15,Wehmeyer19,Thielemann17,Kobayashi20,Wanajo21,Farouqi22,Molero25}.
Thus, the origin of the heavy elements is a fundamental topic of recent investigations (for a review, see \citealt{2021RvMP...93a5002C}).

Light $r$-process elements such as Sr were directly detected in the aftermath of the merging of two neutron stars in gravitational wave event GW170817 (e.g., \citealt{Watson19,Domoto21}), establishing neutron star mergers (NSMs) as primary site of the $r$-process \citep{watson2019a,2014AIPC.1594..266W,2015MNRAS.448..541J,2019Natur.569..241S,2017ARNPS..67..253T}.
The question arises whether NSMs can explain the entire amount of $r$-process elements, or do also other sites exist where these are formed. 
Particularly, extremely metal-poor stars (EMPS, stars with [Fe/H]\footnote{[A/B]$=\log (\text{A}/\text{B})_\text{star}-\log (\text{A}/\text{B})_\odot$}$\leq$-3) in the Galactic halo (which are interpreted as very old stars) already show a large amount of $r$-process elements \citep[e.g.,][]{Roederer10,Lombardo25}. 
Despite these observations, NSMs are very unlikely at lowest metallicities as pointed out by \cite{Argast04,Cescutti15,Wehmeyer15,Wehmeyer19,2022EPJWC.26009002T,2022A&A...663A..70F}, and therefore the investigation of the early formation of heavy elements is an essential, unsolved problem.
As such early-acting source, multiple sites have been investigated:
\begin{itemize}
    \item \textit{Supernova explosions.} Despite earlier investigations finding a high entropy neutrino wind with the right properties (\citealt{Woosley94}, \citealt{Takahashi94}), more advanced CCSN simulations (e.g., \citealt{Liebendorfer03}) led to proton-rich environments in the innermost ejecta (e.g., \citealt{Fischer10}, \citealt{Hudepohl10}), causing the $\nu p$-process (\citealt{Frohlich06a}, \citealt{Frohlich06b}, \citealt{Pruet05}, \citealt{Pruet06},  \citealt{Wanajo06}), and no $r$-process.
    \item \textit{Collapsars.}~Collapsars are supernova explosions with a black hole at their center \citep[c.f.][and references therein]{MacFadyen99,Proga03,Ott11,Aloy21}, which have been proposed as a potential site for the $r$-process \citep[c.f.][]{Fujimoto08,Siegel19}. However, collapsar models are still not self-consistent, rely on an accretion disc around a rotating and magnetized black hole, and only feature incomplete and approximate neutrino transport. Hence, quantifying the impact of such a scenario is still difficult.
    \item \textit{Magneto-rotationally driven supernovae} \citep[MHD-SN, or ``Jet-SNe'', c.f.,][]{Fujimoto06,Fujimoto08,Winteler12} whose rotation and magnetic fields produces a jet along the rotation axis of the supernova, where the conditions for the $r$-process could be met \citep[e.g.,][]{Winteler12,Nishimura17}. Newer simulations have, however, identified a magneto-hydrodynamical instability \citep{Mosta14}, which causes the forming jet to break.
    \item \textit{Neutron star-black hole (NSBH) mergers}, which would have the advantage of increasing the metallicity of the interstellar medium by only one successful CCSN explosion, instead of two (as in the NSM case). Unfortunately, their early contribution is difficult to quantify and relies on a parametrized approach \citep[e.g.,][]{Wehmeyer19}.    
    \item \textit{Primordial black hole (PBH) - millisecond pulsar (MSP) interaction}, where a PBH captured by an MSP sinks to the center of the NS and consumes it from the inside. Because of the resulting spin-up, $\approx$0.1 M$_\odot$–0.5 M$_\odot$ of relatively cold neutron-rich material gets ejected, which is consistent with spatial distributions in the Galaxy and ultrafaint dwarfs, as well as with the $r$-process content and evolution histories in these sites \citep[c.f.][]{Fuller:2017uyd}.
\end{itemize}
It is not clear to which extent these events contribute to the present inventory of heavy elements, and/or whether NSMs alone can explain the $r$-process enrichment of the Milky Way, see, e.g., \cite{2023ApJ...943L..12K}.

An illustration of the low metallicity enrichment problem can be found in Fig.~\ref{EuFeBenjamin} taken originally from \cite{Wehmeyer15} and adapted:

\begin{figure}[t]
  \centering
  \includegraphics[width=0.99\linewidth]{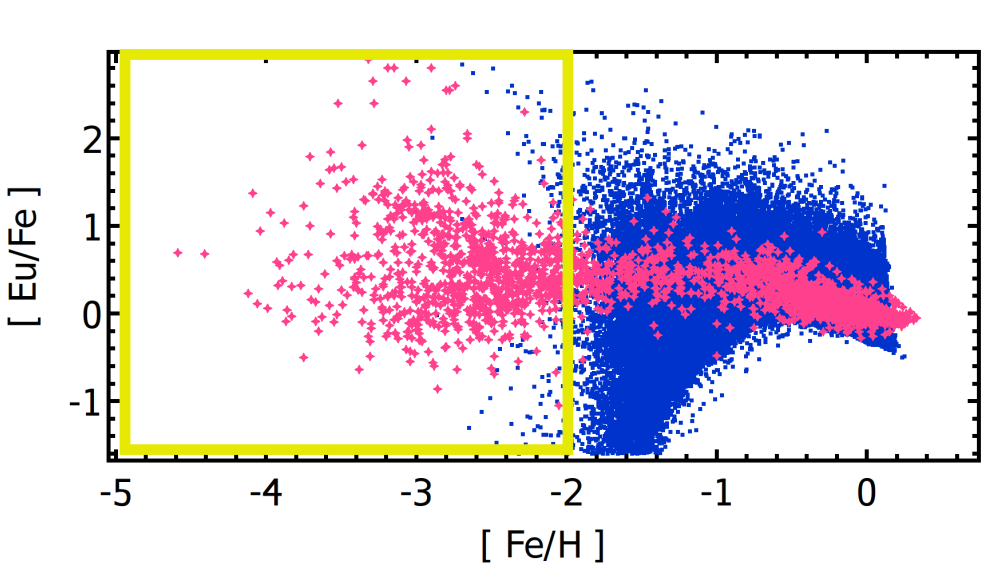}
   \caption{An illustration of the low metallicity issue of the $r$-process. Blue model stars represent a GCE model where NSMs act as \textit{exclusive} $r$-process site. These start to contribute to the Galactic $r$-process inventory only at [Fe/H] $> -2$, whereas EMPS with very high $r$-process abundances have been observed (magenta stars represent observations, the yellow box denotes the area of interest of the low metallicity issue of the $r$-process). Figure taken from \cite{Wehmeyer15} and adapted. See Data Availability Section for data sources.} 
   \label{EuFeBenjamin}
   
\end{figure}

As europium (Eu) is almost exclusively made by the $r$-process, the  [Eu/Fe] versus [Fe/H] relation can be used to test whether a proposed astrophysical process can describe the chemical evolution of our Galaxy.
Whereas a close correlation is seen for stars with [Fe/H] $> -2$, a large scatter is observed in the range [Fe/H] $< -2$, which is difficult to reconcile with NSMs acting as \textit{exclusive} $r$-process site (e.g., \citealt[e.g.,][]{Cescutti15,Wehmeyer15,Wehmeyer19,Thielemann17,Kobayashi20,Farouqi22}).

In this work, we will not present a solution of the question which process is responsible for the production of the heavy elements. 
Instead, we present a phenomenological approach. Having noticed that the deficiency of $r$-process elements mainly concerns very- and extremely metal poor stars, we consider pre-enrichment. To study the effects of such pre-enrichment, we set up a GCE model with pre-enriched infall to determine whether such infall would cure the issues of the above-mentioned sites, and be able to support the $r$-process abundances in the very metal poor star regime and below.

We then speculate about a potential origin of this pre-enriched infall: We noticed that all of the early-acting sources discussed above exhibit a hot, dense, and neutron-rich environment.
As the matter expands, the equilibrium composition freezes, and the subsequent evolution of the composition is described by reaction kinetics. In our approach, we assume that even in the early Universe, there is some hot and dense matter (referred to here as pre-enriched infall) that already contains a considerable amount of heavy elements.
We extract the properties of that source at freeze-out from the data, but we do not give a microscopic approach which describes the dynamical process how these freeze-out states can occur. Instead, we infer from the observed heavy element distribution parameter values which can characterize the variability in the heavy-element distribution, and consider this as a signature of the scatter of local conditions in the early Universe.

To characterize the properties of such an additional source, we consider the heavy-element freeze out (HEFO) concept given in \citet{2025Univ...11..323R} and use this as an early-acting, primordial nucleosynthesis-era acting heavy element production site.
We consider values for the Lagrange parameters, the temperature and chemical potentials at freeze-out, to infer the observed data.
in particular, we are able to describe the large [Eu/Fe] values at low metallicities.  
The HEFO may serve as initial condition to run a nuclear reaction network code, for instance SkyNet \citep{Lippuner:2017tyn} or WinNet \citep{reichert2023a}, to obtain the detailed final isotopic abundances. 
Unlike standard approaches, which assume a nuclear statistical equilibrium (NSE) as an initial condition at temperatures of $T\sim 0.5$ MeV,
the distribution of heavy elements freezes out even at higher temperatures and densities as soon as the neutron-rich environment disappears.

This work is organized as follows:
After introducing the chemical evolution model in Section~\ref{sec:model}, we use its results (Section~\ref{sec:GCEResults}) to make assumptions about the possible contribution parameters of an early acting, spatially variable heavy element source. In Section~\ref{sec:Results}, we suggest an interpretation of the results of the GCE model.
Finally, we draw conclusions, and discuss our findings in Section~\ref{conclusion}.

\section{The GCE model} \label{sec:model}

Unlike homogeneous Galactic chemical evolution models, inhomogeneous approaches explicitly follow the spatial distribution of nucleosynthesis events. This allows them to reproduce the observed scatter in elemental abundances rather than only their mean evolutionary trends. A limitation of this approach is that large-scale Galactic processes, such as spiral-arm mixing or galaxy interactions, can only be represented approximately.

In this work, we employ the inhomogeneous chemical evolution model introduced by \cite{Wehmeyer15,Wehmeyer19}. The following subsections briefly summarize the main ingredients of the model for completeness (Sections~\ref{model:setup} and \ref{model:nucleosynthesissites}) and describe the modifications implemented for the present study, with particular emphasis on the inclusion of an additional $r$-process production channel associated with primordial infall (Section~\ref{model:Infall}).

\subsection{General setup}\label{model:setup}

The simulation considers a cubic Galactic volume of $(2\,\mathrm{kpc})^3$, subdivided into $40^3$ computational cells, each with a side length of 50 pc.

At every simulation step of 1 Myr, the following sequence of calculations is carried out:

\begin{enumerate}

\item Primordial gas is assumed to accrete uniformly into every simulation cell. The total gas infall into the computational domain follows the prescription

\begin{equation}
\dot{M}(t)=a\,t^b\,e^{-t/\tau}\textit{,}
\end{equation}

which describes an initial increase in the accretion rate followed by an exponential decline. The parameters $\tau$ and the total simulation time $t_{\mathrm{final}}$ are fixed at the beginning of the calculation. The remaining parameters, $a$ and $b$, are determined from the total accreted mass,

\begin{equation}
M_\mathrm{tot}=\int_0^{t_{\mathrm{final}}}a\,t^b\,e^{-t/\tau}\,dt\textit{,}
\end{equation}

together with the time at which the infall rate reaches its maximum,

\begin{equation}
t_\mathrm{max}=b\tau\textit{.}
\end{equation}

The adopted infall parameters are summarized in Table~\ref{infall parameters}.
\end{enumerate}

\begin{table}[htp]
\begin{center}
\begin{tabular}{|llr|}
\hline
\hline
$M_{\rm tot}$ & total infall mass & $10^8 \text{M}_{\odot}$ \\
$\tau$ & time scale of infall decline & $5\times10^9$ yrs \\
$t_{\rm max}$ & time of the highest infall rate & $2\times10^9$ yrs \\
$t_{\rm final}$ & duration of the simulation & $13.6\times10^9$ yrs \\
\hline
\end{tabular}
\caption{Main infall parameters. See Wehmeyer et al. (2015) for a detailed discussion of these parameters.}
\label{infall parameters}
\end{center}
\end{table}

\begin{enumerate}[resume]
\item The total gas mass within the simulation volume is then evaluated to determine the star formation activity. We adopt a Schmidt law with a density exponent of $\alpha=1.5$, motivated by the expectation that star formation depends both on the local ISM density and on cloud--cloud interactions (see \citealt{Schmidt59,Kennicutt98,Larson91}). The resulting stellar mass formed during the current time step is converted into the corresponding number of newly formed stars by integrating a Salpeter-type initial mass function (IMF) with slope $-2.35$.
\item After determining the number of stars formed, star-forming cells are selected randomly. Since processes such as cloud--cloud collisions or expanding supernova shells are more likely to trigger star formation in dense environments, cells with higher gas densities are preferentially chosen.

\item Once a star-forming cell has been selected, the mass of each newly formed star is sampled randomly according to a Salpeter-type IMF with slope $-2.35$ over the mass interval $0.1\,\text{M}_{\odot}\leq m\leq50\,\text{M}_{\odot}$\footnote{Throughout this manuscript, stellar masses (excluding compact remnants such as NSs and BHs) refer to the zero-age main-sequence mass.}. To avoid an unrealistically bottom-heavy IMF, star formation is permitted only in cells containing at least $50\,\text{M}_{\odot}$ of gas. Stars with initial masses below $10\,\text{M}_{\odot}$ are classified as low- and intermediate-mass stars (LIMS), whereas more massive stars are treated as high-mass stars (HMS).

\item Each newly formed star inherits the chemical composition of the ISM from which it forms. Its lifetime is determined from its birth mass and metallicity using the stellar evolution calculations of the Geneva Stellar Evolution and Nucleosynthesis Group (cf. \citealt{Schaller92,Schaerer93a,Schaerer93b,Charbonnel93}):

\begin{equation}
\begin{split}
\log(t)&=(3.79+0.24Z)-(3.10+0.35Z)\log(M)\\
&+(0.74+0.11Z)\log^2(M)\textit{,}
\end{split}
\end{equation}

where $t$ denotes the stellar lifetime in Myr, $Z$ is the metallicity relative to solar, and $M$ is the stellar mass expressed in solar masses.

\item Once the calculated stellar lifetime has elapsed, the corresponding stellar death event is initiated according to the star's initial mass, as described in the following subsections.

\end{enumerate}

\subsection{Nucleosynthesis sites}\label{model:nucleosynthesissites}

\subsubsection{Low and intermediate mass stars}\label{model:LIMS}

Low- and intermediate-mass stars (LIMSs) are responsible for producing the majority of carbon and nitrogen in the Galactic disk \citep[e.g.,][]{Kobayashi11}. During their Asymptotic Giant Branch (AGB) evolution, these stars also synthesize most of the heavy nuclei beyond Sr through the slow neutron-capture process (the ``$s$-process'') that are observed in the Solar System \citep[e.g.,][and references therein]{Kappeler11,Bisterzo14}.

In contrast, LIMSs contribute only negligibly to the Galactic inventories of Fe and Eu. Consequently, within our simulation they are treated primarily as reservoirs that temporarily lock up interstellar gas during their lifetimes. Although these stars eventually return substantial amounts of H and He to the ISM, thereby producing a small dilution of the local [Fe/H] abundance, this effect has no significant impact on the results or conclusions presented here.

At the end of their evolution, LIMSs lose part of their mass through stellar winds, forming planetary nebulae and leaving behind white dwarfs. Observed planetary nebulae typically extend over only a few tenths to a few light years (e.g., the Cat's Eye Nebula NGC 6543 has a diameter of about 0.2 light years, \citealt{Reed99}, while the Helix Nebula spans approximately 2.87 light years, \citealt{Odell04}). These scales are smaller than the spatial resolution of our model, and therefore the expelled material is simply returned to the simulation cell in which the progenitor star resides once its lifetime has ended.

\subsubsection{Thermonuclear supernovae}\label{model:SNIa}

Since a significant fraction of stars forms in binary systems (e.g., \citealt{Duchene13}), some newly born stars will eventually evolve into thermonuclear supernovae of type Ia (SNIa). Following the analytical prescription of \cite{Greggio05}, we avoid explicitly modelling the complex evolution of binary systems and instead assign each newly formed intermediate-mass star (IMS, $1\,\mathrm{M}_\odot \leq m \leq 10\,\mathrm{M}_\odot$) a probability of $P_{\mathrm{SNIa}}=9\times10^{-4}$ of belonging to a binary that will ultimately produce a SNIa event.

This prescription corresponds to a SNIa occurrence rate of $7.49\times10^{-4}$ events per solar mass of stars formed. When the second IMS in such a binary reaches the end of its lifetime, we inject $10^{51}$ erg of energy into the surrounding ISM together with the appropriate nucleosynthetic yields (cf. \citealt{Iwamoto99}, model CDD2). As in \cite{Wehmeyer15}, we assume that each SNIa ejects the same amount of Fe regardless of metallicity. Although this simplification is unlikely to be fully realistic (e.g., \citealt{Timmes03,Thielemann04,Travaglio05,Bravo10,Seitenzahl13,Leung18}), it has only a minor influence on the predictions of our model.

While SNIa events do not synthesize $r$-process material, they represent the dominant source of iron enrichment in the Galactic disk \citep[e.g.,][]{Matteucci86}. Their contribution is therefore essential for reproducing the observed evolution of the Galactic [Eu/Fe] abundance ratio.

\subsubsection{Core collapse supernovae}\label{model:SNe}

Stars with initial masses above $10\,\mathrm{M}_{\odot}$ proceed through all advanced burning stages up to silicon burning, ultimately forming an iron core \citep[e.g.,][]{Jones13}. 
Once nuclear energy generation in the core ceases, gravity overwhelms the pressure support and the stellar core undergoes collapse. Compression continues until nuclear densities are reached, forming a proto-neutron star (proto-NS). 
Neutrino and antineutrino absorption in the gain region deposits energy into the surrounding matter (e.g., \citealt{Burrows13,Janka16,Janka17,Burrows18}), and if this heating is sufficiently efficient, a successful core-collapse supernova (CCSN) results. 
In our code, CCSNe eject yields acording to \cite{Thielemann96,Nomoto97}, deposit 1 Bethe of kinetic energy into the local simulation volume, and leave behind a NS.

\subsubsection{Neutron star mergers} \label{subsec:NSM}

When both members of a binary system are high-mass stars, each ultimately ends its evolution either as a successful CCSN or as a failed supernova \citep[e.g.,][]{Nomoto13}. If the compact remnants remain gravitationally bound despite the natal kicks associated with the explosions (e.g., \citealt{Tauris17}), the system gradually loses orbital energy through the emission of gravitational waves and eventually merges as a neutron star merger (NSM) in the case of two successful CCSN explosions.

Predicted NSM rates span a wide range (e.g., \citealt{Kalogera04}), while the occurrence frequency of BHNSMs remains even more uncertain (e.g., \citealt{Mennekens14}). Likewise, both nucleosynthesis studies (e.g., \citealt{Abbott1,Chornock17,Cowperthwaite17,Kasen17,Tanaka17,Wang17,Gompertz18,Hotokezaka18,Rosswog18}) and Galactic chemical evolution calculations (e.g., \citealt{Matteucci14,Cescutti15,Hirai15,Ishimaru15,Shen15,Wehmeyer15,Komiya16,Haynes19}) adopt substantially different merger rates. A compilation by \cite{Cote17} demonstrated that assumed NSM frequencies differ by roughly two orders of magnitude among modern GCE models. Much of this variation originates from differences in the adopted infall prescriptions, star formation models, initial mass functions, CCSN and SNIa yields, and the assumed ejecta masses per merger event. After normalizing these quantities to a common IMF and consistent Fe and Eu yields, however, the inferred NSM frequencies converge to within approximately a factor of four \citep{Cote17}.

Although theoretical predictions for merger rates remain uncertain, direct detections of gravitational waves now provide valuable observational constraints on compact binary merger frequencies. While the earliest detections corresponded to BH--BH mergers, later observations identified a neutron-star merger (e.g., \citealt{Abbott4,Abbott5}), implying a NSM rate of $1540^{+3200}_{-1220}\,\mathrm{Gpc}^{-3}\,\mathrm{yr}^{-1}$.

To limit the number of free parameters in our model, we adopt a simplified prescription based on an \textit{effective} probability, $P_{\mathrm{r-proc}}$, representing the likelihood that a newly formed high-mass star belongs to a binary system that will eventually produce an $r$-process event through either a NSM. Throughout this work we assume $P_{\mathrm{r-proc}}=4\%$. Under the assumptions of an integrated Salpeter IMF with slope $-1.35$, a standard cosmic star formation history, and constant compact-binary merger delay times (see \citealt{Cote17} for details), this probability corresponds to $1.03\times10^{-4}$ merger events per solar mass of stars formed. Although this value is somewhat higher than those adopted in many previous GCE studies (see \citealt{Cote17} for a comparison), it translates into an event rate of approximately $1800\,\mathrm{Gpc}^{-3}\,\mathrm{yr}^{-1}$, which remains just slightly above the rates inferred from LIGO/Virgo observations.

In our simulation, every NSM produces $8.5 \cdot 10^{-5} \mathrm{M}_\odot$ of Eu \citep[e.g.,][]{Freiburghaus99,Panov08,Korobkin12,Bauswein13,Rosswog13a,Rosswog13b,Rosswog14,Wanajo14,Eichler15,Just15,Vassh19}
\subsubsection{Primordial heavy element infall}\label{model:Infall}
With a possible interpretation of primordial fluctuations leading to HEFO in mind, we add pre-enriched infall to sub-cells of the simulations. Since cosmological fluctuations are rare and do not occur uniformly throughout space, but are spread far away from each other, we set up four different models to account for the spatial separation of the occurrence of such pre-enriched infall: In model 1 (2, 3, 4) we add [Eu/Fe]=$+3$ (and/at [Fe/H]=$-7$, to aim for a contribution of [Eu/Fe]=$+1$ at [Fe/H]=$-5$) material in every second (third, fifth, eigth) sub-cell of the simulation volume. This corresponds to pre-enriched infall occurring with a spatial separation of 100 (150, 250, 400)~pc. This pre-enriched material mixes with the already existing material, so the contributed effect of this pre-enriched infall becomes smaller as time progresses.

\section{Results of the heavy infall GCE modeling}\label{sec:GCEResults}
The results of using $r$-process element enriched infall in our GCE model can be seen in Figure~\ref{Fig:Res1}. The models with 100pc and 150pc distance between $r$-process enriched infall sub-cells (top panels in Figure~\ref{Fig:Res1}) overpredict Eu abundances almost over the entire metallcitiy range. This happens, because there is not enough distance between Eu-poor sub-cells and sub-cells with fresh Eu falling in during every time step. This makes it difficult to reach much lower abundance values in non-$r$-process infall sub-cells. Only over time, as CCSNe contribute more Fe to the overall simulation volume, the [Eu/Fe] value decreases, and reaches Solar values at Solar metallicities. 
The opposite case can be seen in the model with 400 pc distance between $r$-process enriched infall sub-cells (bottom right panel in Figure~\ref{Fig:Res1}): Here, the $r$-process enriched infall sources are too far apart to be able to provide a sufficient amount of $r$-process material to the overall simulation volume, so abundances stay low. In other words: Any newly added Eu is immediately pushed around in the simulation volume by following CCSN events, so that the high abundance in the enriching sub-cell cannot be sustained, i.e., the high Eu abundance gets diluted very efficiently, leading to lower abundances. Another interesting fact can be witnessed here rather clearly: The infall-supported ``higher'' (i.e., above the floor) Eu abundances (but still too low to explain the observations) between -3$\leq$[Fe/H]$\leq$-2 give way to the abundance ratio dictated by NSMs at around [Fe/H]=-1.3.

We find that a good distance between $r$-process element enriching sub-cells seems to be 250~pc: Here, the distance between infall-enriching sub-cells and those that receive subsequently propagating enriched material seems to be just right to reproduce the observed $r$-process abundances at all metallicities.

The picture becomes even clearer when comparing abundance histograms at  of the four models with observations in the [Fe/H]$\leq$-3 regime only: In Figure~\ref{Fig:Histograms}, the models with enrichment distances of 100 pc (red line) and 150 pc (blue line) overpredict the observed abundances (black line) in the high abundance ([Eu/Fe]$\geq$+1.5) area, whereas the 400 pc enrichment distance model (purple line) underpredicts the observed abundances (black line) in the medium and higher abundance area ([Eu/Fe]$\geq$+0.5) area. The only model with a relatively good agreement to the observed abundances (black line) in that metallicity range seems to be the 250 pc enrichment distance model (blue line), with a good agreement in the mid- to higher abundance range ([Eu/Fe]$\geq$+0.25). This model, however, features too many model stars in the lower abundance range ([Eu/Fe]$\leq$+0.25), where it rather strongly deviates from the observed abundances (black line). This discrepancy, however, could originate in a technological bias in the observations: When analyzing abunances in EMPSs, the Fe lines become fainter with lower [Fe/H]. On the other hand, also the Eu lines in that metallicity range become fainter, which could potentially lead to less detections in the low [Fe/H], low [Eu/Fe] area, which could hide a whole population of stars which feature undetectably faint Eu lines. This makes it difficult to establish a histogram of [Eu/Fe] in the range [Eu/Fe]$\lessapprox$0 as a reliable benchmark for a model which predicts stars in this metallcitiy range.

\begin{figure*}[htbp]
\begin{center}
\includegraphics[width=0.45\linewidth]{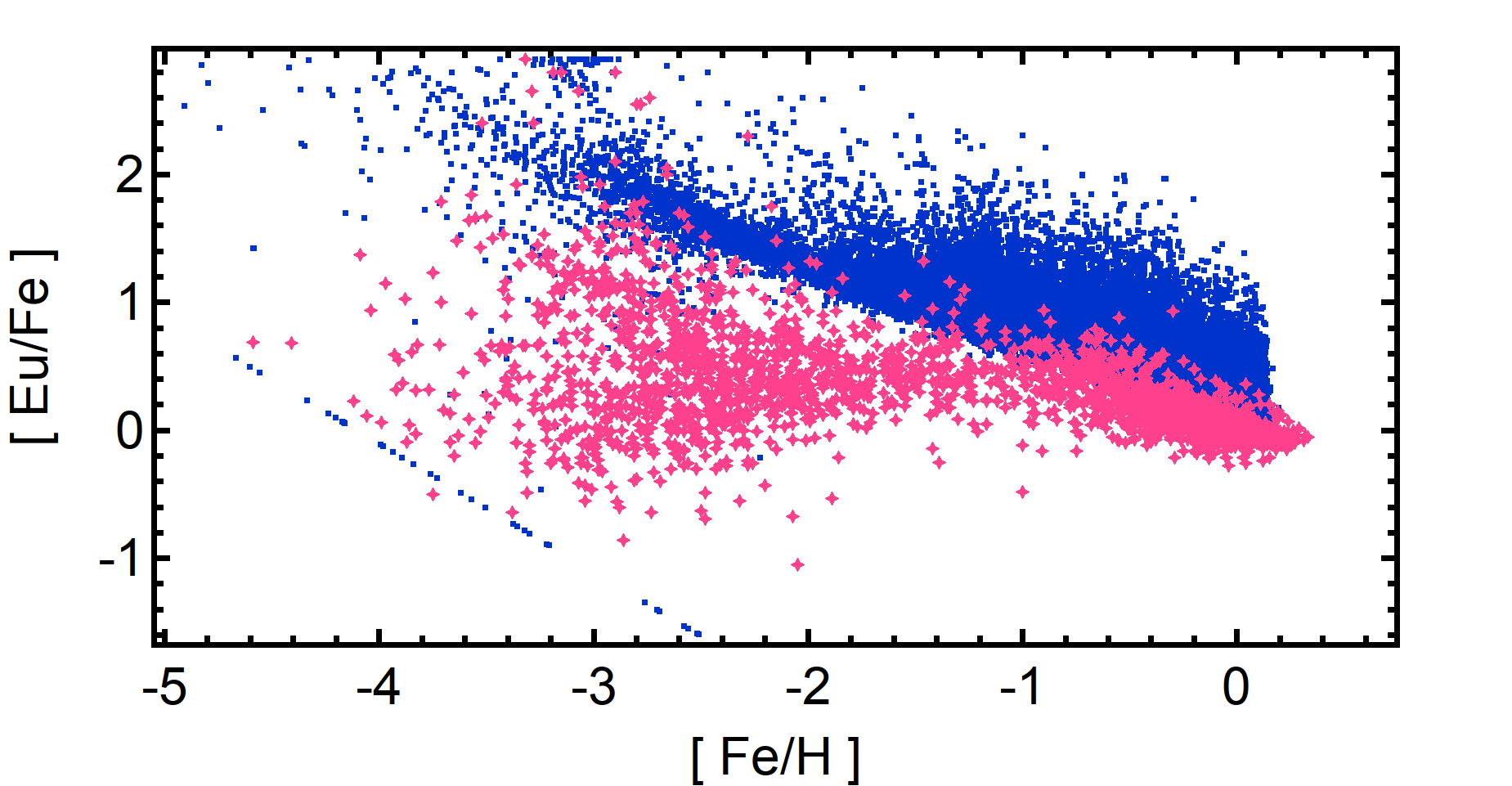}
\includegraphics[width=0.45\linewidth]{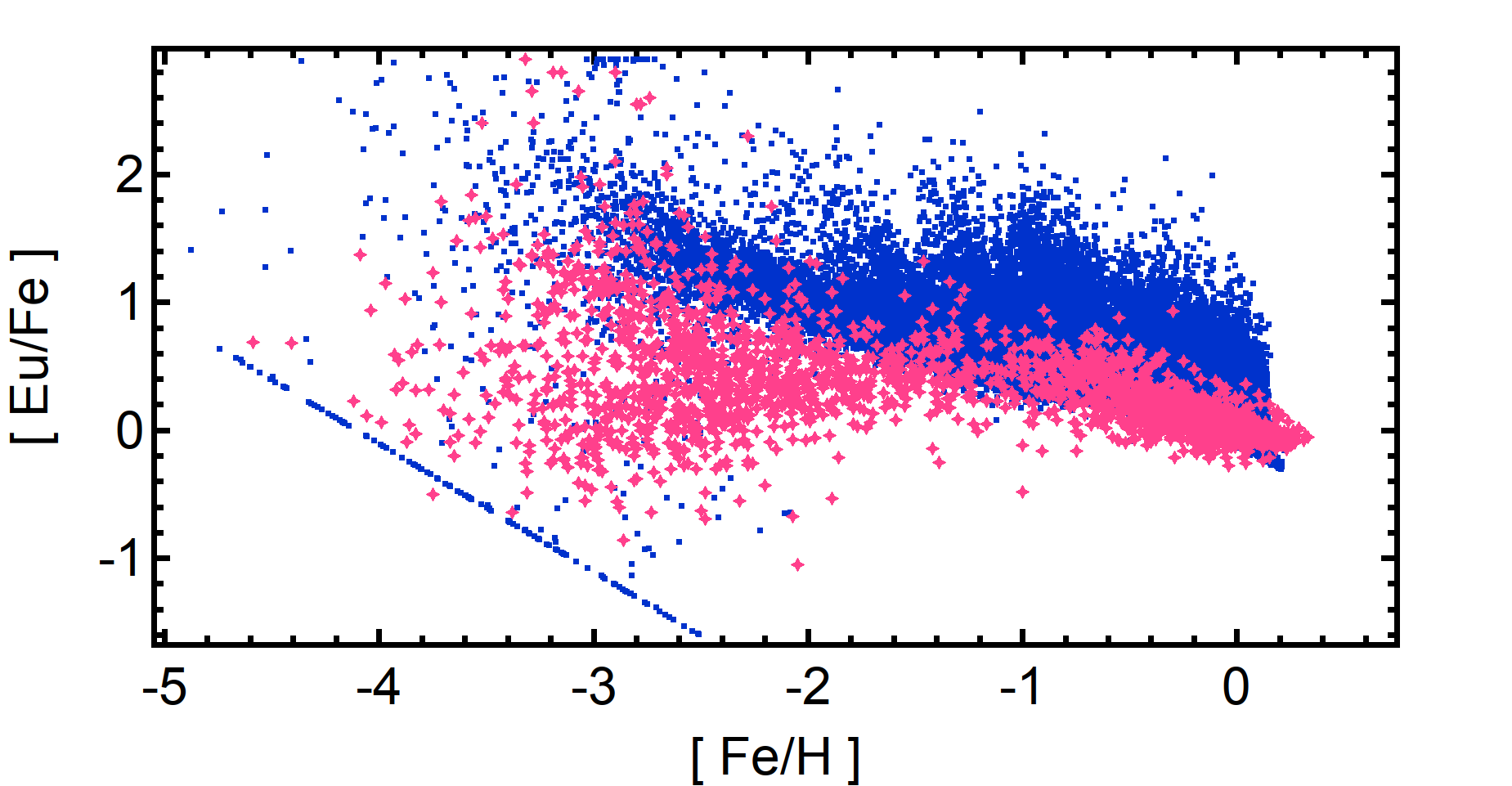}
\includegraphics[width=0.45\linewidth]{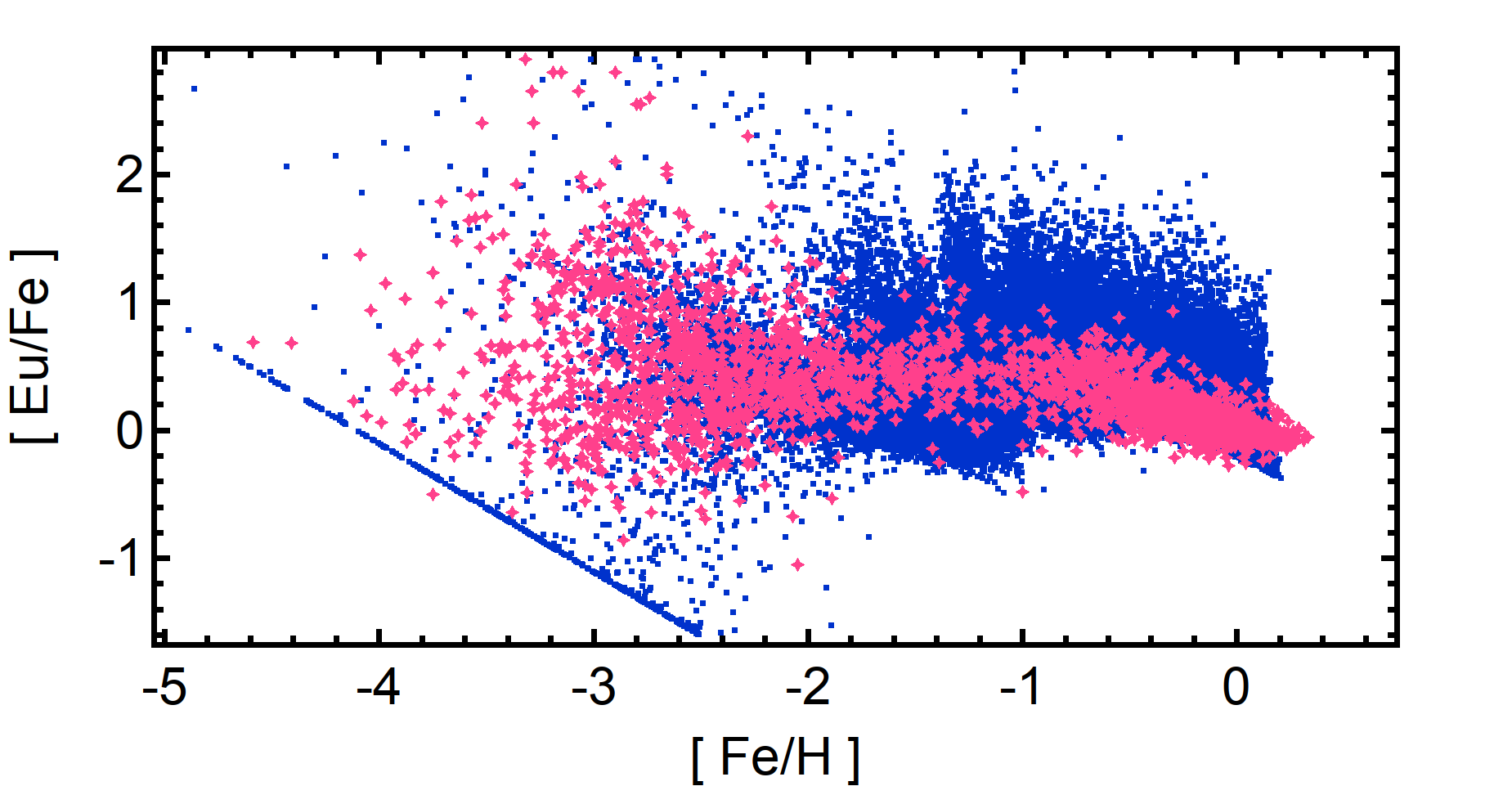}
\includegraphics[width=0.45\linewidth]{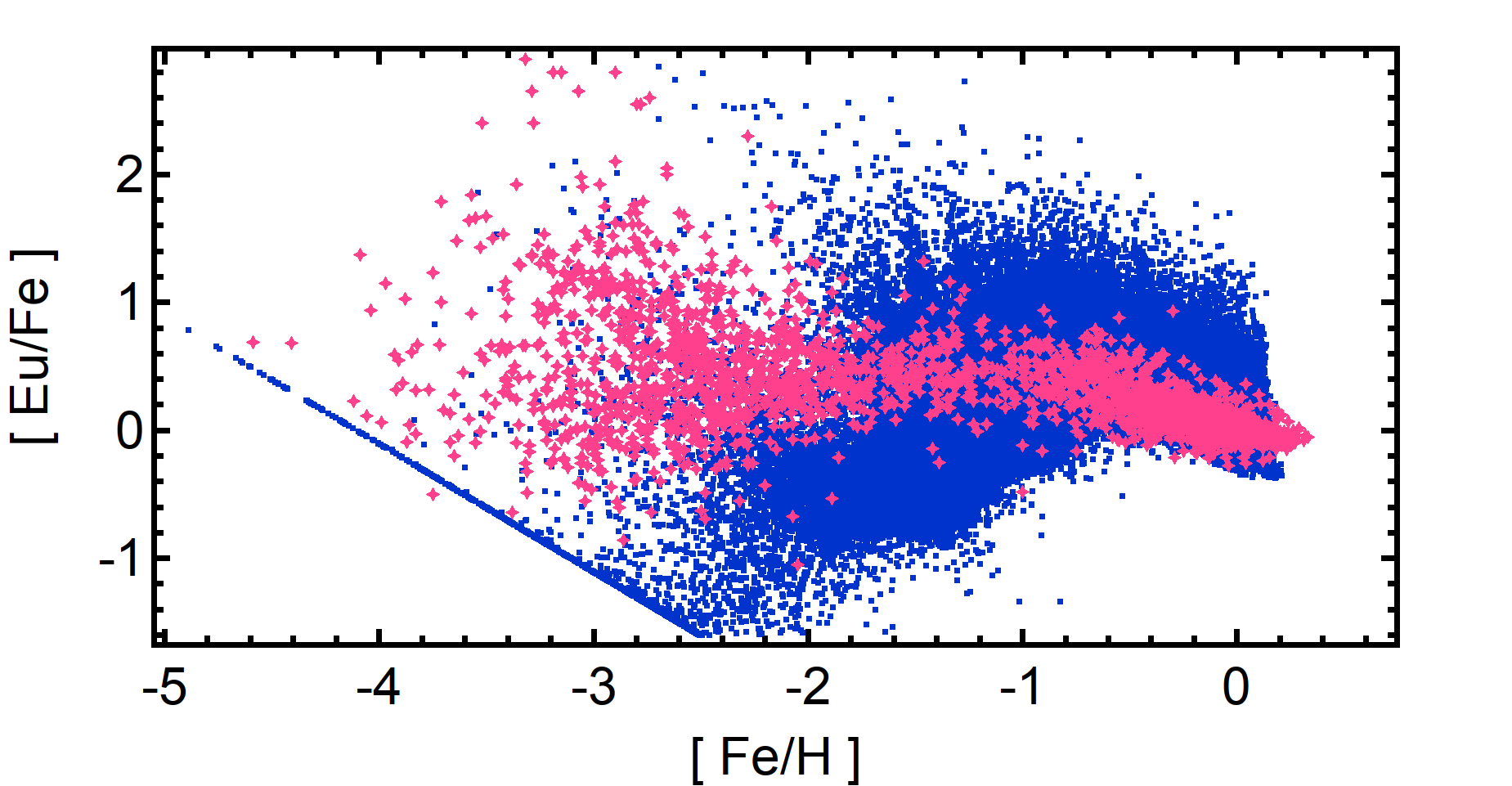}

\caption{Results of the inhomogneneous infall simulations. Different panels represent different distances (top left: model~1, 100~pc, top right: model~2, 150~pc, bottom left: model~3, 250~pc, and bottom right: model~4, 400~pc) between sub-cells with heavy element enriched infall. Magenta stars represent observations, whereas blue dots represent model stars. 
(Note the numerical artifact of a diagonal downward moving line, representing the ``indisturbed'' infall curve, which can be interpreted as a floor.)}
\label{Fig:Res1}
\end{center}
\end{figure*}

\begin{figure}[htbp]
\begin{center}
\includegraphics[width= \linewidth]{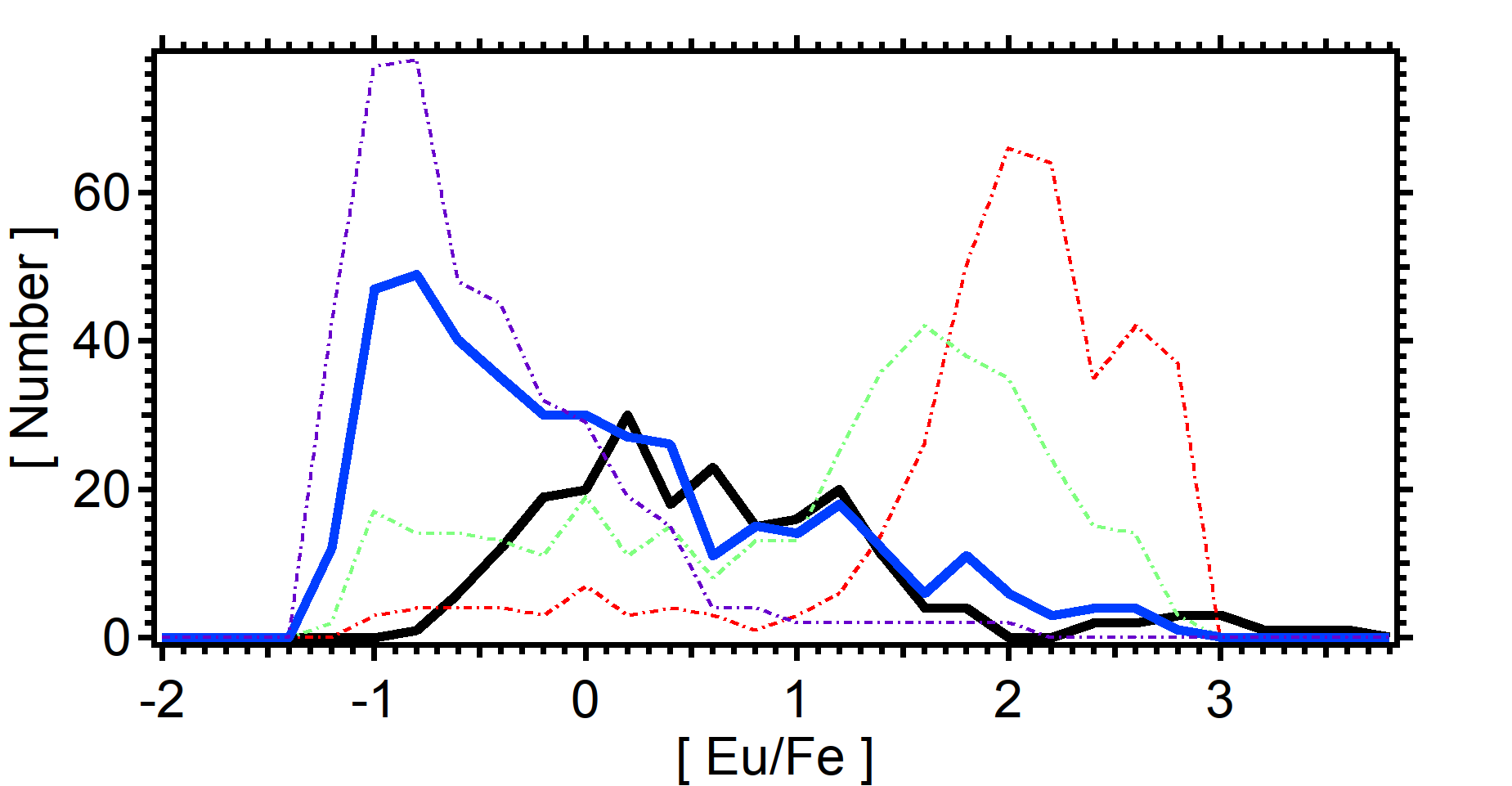}

\caption{[Eu/Fe] histogram (bin width $0.2$) of all stars below [Fe/H]$\leq$-3 in Fig.~\ref{Fig:Res1}.\\
Black: observations.
Red (dash-dotted line): Model~1 (100~pc between enrichment locations).
Green (dash-dotted line): Model~2 (150~pc between enrichment locations).
Blue: Model~3 (250~pc between enrichment locations).
Purple (dash-dotted line): Model~4 (400~pc between enrichment locations).
}
\label{Fig:Histograms}
\end{center}
\end{figure}

\section{Interpretation of the results} \label{sec:Results}
\subsection{The origin of the heavy elements in the early Universe}
\label{sec:Cosmology}
As shown in the previous Section, the observed high abundances of $r$-process elements at low metallicities can be explained by a GCE model featuring pre-enriched infall. In the following, we speculate about a possible early production channel which might act as such a source.

The problem of the formation of heavy elements in the early Universe arises from the fact that, under the scenario of homogeneous Big Bang nucleosynthesis (HBBN), only the lightest elements are produced.
An interesting alternative is inhomogeneous Big Bang nucleosynthesis (IBBN); see \citet{Gonin:2025uvc}.
It is assumed that primordial black holes and other large-scale density fluctuations already existed before Big Bang nucleosynthesis. Such density fluctuations, which survive the timescale of HBBN, are potential sites where the nucleosynthesis of heavy elements can take place.

In this paper, we do not address the sites where the $r$-process elements were formed in the early Universe, we rather discuss the conditions necessary to produce the observed [Eu/Fe] ratios.
This is a phenomenological approach; the determination of astrophysical scenarios and possible sites of heavy-element formation is the subject of future work.

\subsection{The HEFO concept}
\label{distributions}

The formation of the $r$-process elements requires specific conditions: a neutron-rich, hot, and dense environment.
We describe the decompression and cooling of hot and dense nuclear matter within the framework of a freeze-out approach.
Detailed balance calculations lead the system to a local thermodynamic equilibrium.
If certain reactions proceed slowly compared to the relative rate of change of the thermodynamic parameters, the corresponding degrees of freedom “freeze out,” and their average values determine the subsequent behavior of the system. These degrees of freedom must be included in the set of relevant observables that characterize the non-equilibrium state of the system; see \citep{Zubarev}.

The evolution of the chemical composition following the “freeze-out” is described by nuclear reaction network (NRN) codes
such as SkyNet \citep{Lippuner:2017tyn} and WinNet \citep{reichert2023a}.
The initial conditions, which are determined by the thermodynamic parameters of temperature and chemical potentials during the “heavy element freeze-out” (HEFO), are regarded as phenomenological parameters that characterize the conditions under which the $r$-process elements are formed.

We are interested in the composition of matter, expressed by the mass fraction $X_{AZ}=A n_{AZ}/n_B$, where $A$ and $Z$ are the mass numbers and charge numbers, respectively, of the isotopes with density $n_{AZ}$;
$n_B$ is the baryon number density.
The standard approach to describing the temporal evolution of the composition of matter $X_{AZ}(t)$
for expanding, hot, and dense matter—such as that found in supernova explosions or mergers of compact objects—assumes that a nuclear statistical equilibrium (NSE) exists above a critical temperature $T_{\rm NSE}$.
The critical temperature  $T_{\rm NSE}$ is taken in the range 0.5 - 0.7 MeV,  see, e.g., \citet{ricigliano2024} or \citet{2017PASA...34...67F}. 

This simple approach to describing the non-equilibrium expansion process has several shortcomings.
(i) Standard NRN codes do not account for in-medium effects on binding energies and reaction rates; instead, they assume the free values,
and (ii) not all degrees of freedom freeze under the same conditions.
For example, the production of heavy elements requires a neutron-rich environment,
and the formation of heavier elements ceases when the neutron-rich environment disappears.
The parameter values for HEFO are higher temperatures $T \approx 5$ MeV and baryon number densities $n_B \approx 0.01$ fm$^{-3}$ \citep{2025Univ...11..323R}.

Following this heavy-element freeze-out, $X_A(t)=\sum_Z X_{AZ}(t)$ changes due to the emission and absorption of nucleons, $\alpha$ decay, and fission processes. For the light elements, fusion processes are also possible.
We assume that some general features of the distribution function emerge very early in the chemical evolution process.
Finer details, such as the staggering with respect to $N$ or $Z$, emerge only at a later stage and are described by detailed nuclear reaction networks.

To identify these gross structures of the distribution function, we consider a coarse-grained distribution, the accumulated mass fraction 
\begin{equation}\label{XA}
\hat X_{\hat A}=\frac{1}{n_B} \sum_{A'=\hat A}^{\hat A+3} A' \sum_{Z,\nu} n_{A'Z\nu}
\end{equation}
with $\nu$ denoting the excitation state of the isotope $\{A,Z\}$.
The solar abundances are considered in \cite{2025Univ...11..323R}. 
There, the distributions of the heavy elements were used to infer the Lagrangian parameters $\lambda_T, \lambda_n, \lambda_p$, which describe the nuclear matter at HEFO.
 The parameter values $\lambda^\odot_T=5.266\,\mathrm{MeV}$,
$\lambda^\odot_n=940.317\,\mathrm{MeV}$, and  $\lambda^\odot_p=845.069\,\mathrm{MeV}$ were determined.
The corresponding coarse-grained distribution is shown in Fig. \ref{fig:XA1}.

\begin{figure}
     \begin{center}
 \includegraphics[width=\linewidth]{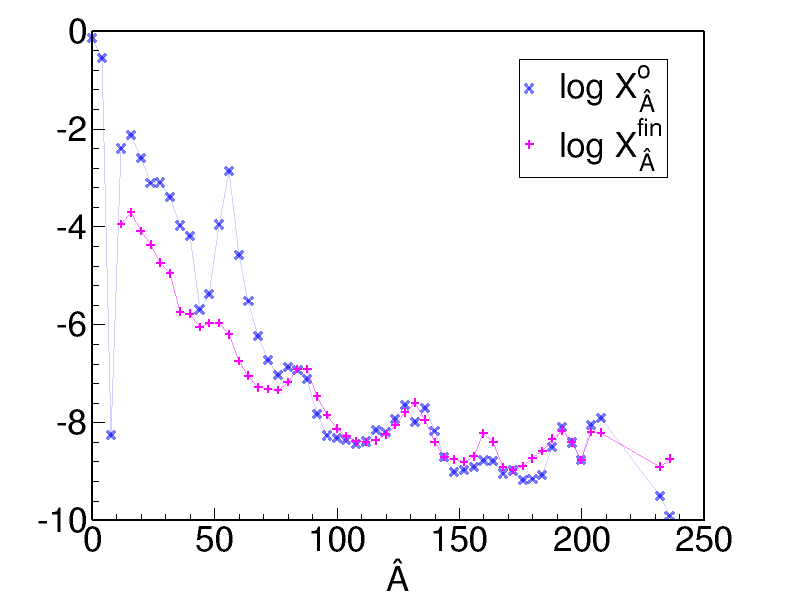}
     \end{center}
     \caption{Accumulated mass fraction $\hat X_{\hat A}$ (red ``$+$''
symbols) for the parameter values $T=5.266\,\mathrm{MeV}$,
$\mu_n=940.317\,\mathrm{MeV}$ and  $\mu_p=845.069\,\mathrm{MeV}$ after
evaporation of neutrons $\hat X'_{\hat A}$ compared with the solar
accumulated mass fraction $\hat X^\odot_{\hat A}$ (blue ``$\times$''
symbols. 
In addition to neutron evaporation, leading to  $\hat X'_{\hat A}$, nuclei
with $A>212$ are subject to $\alpha$ decay (feeding the region near
$A\sim 200$) and fission (feeding the region near $A\sim 160$). See \citet{2025Univ...11..323R} for details.
}
     \label{fig:XA1}
\end{figure}

As shown in Fig. \ref{fig:XA1}, the HEFO model closely reproduces the solar heavy-element distribution,
while light elements are underproduced. In particular, the iron peak is underproduced by a factor of $10^3$.
As already emphasized above, the freeze-out applies only to the heavy elements, while the non-equilibrium evolution of the light elements continues, particularly burning processes that can increase the abundances of the light elements.
  The Lagrange parameters that describe the HEFO model are not universal.
Deviations from the universality of the heavy-element distribution can be described phenomenologically by the appropriate choice of Lagrange parameters, as shown in \cite{2025FrASS..1233496B}.

Assuming that low-metallicity stars are old stars, the high [Eu/Fe] values at [Fe/H] $\le$ -2 pose a problem for standard models, according to which no heavy elements were present in the very early Universe.
Furthermore, in the low-metallicity range [Fe/H] $\le$ -2, we do not observe a close correlation between [Eu/Fe] and the “time scale” [Fe/H], but rather a high degree of dispersion.
From a phenomenological perspective, the observed large fluctuations in Eu abundances at low metallicities can be explained by local inhomogeneities \citep{Wehmeyer:2015sra}.
A homogeneous environment does not emerge until a later stage in the GCE as a result of mixing.

To discuss the evolution over time of the relative abundance [Eu/Fe] of an individual star, we consider
nuclear fusion processes that have been occurring continuously in the Universe since the earliest times, as soon as stars came into existence. The total amount of Fe increases, but the abundance of the heavy element Eu remains unchanged.
The abundance of H also does not change significantly.
This results in a linear relationship:
\begin{equation}
\label{linFe}
    [{\rm Eu}/{\rm Fe}(t)]=[{\rm Eu}/{\rm Fe}(t_0)]-[{\rm Fe}(t)/{\rm H}]+[{\rm Fe}(t_0)/{\rm H}].
\end{equation}
This relationship holds as long as no Eu (being an $r$-process element) is produced.

To discuss the region of “old stars” with $[{\rm Fe}/{\rm H}] < -2$, we note that we find high values up to $[{\rm Eu}/{\rm Fe}] > 2$, which exhibit significant dispersion.
Using our phenomenological approach we can describe the conditions for the formation of these heavy elements.
Within the framework of the HEFO concept, we introduce the Lagrange parameters $\{\lambda_T,\lambda_n,\lambda_p\}$.
These parameter values are fitted to the abundances of the heavy elements; for example, for the solar distribution we use the values $\lambda^\odot_i$ given above, see  Fig. \ref{fig:XA1}, where all elements are shown.
The solar abundances of the light elements up to Fe are underproduced at the time of the HEFO, as subsequent fusion processes can alter the abundances of the light elements.
In particular, Fe is underproduced by a factor of $10^3$,
 $[{\rm Fe}/{\rm H}]^{\rm HEFO} \approx -3$, whilst the solar abundance of Eu is approximately reproduced, such that  $ [{\rm Eu}/{\rm Fe}]^{\rm HEFO} \approx 3$.
This example shows that, in the HEFO scenario, distributions with low metallicity but a high value of $ [{\rm Eu}/{\rm Fe}]$ can occur.

Next, we discuss the scatter of [Eu/Fe] among stars with low metallicity [Fe/H] $<$ -2 (we consider the ‘yellow box’ of stars with [Fe/H] $<-2$, see Fig. \ref{EuFeBenjamin}).
As shown in Ref. \cite{2025FrASS..1233496B}, different stars are characterised by different sets of Lagrange parameters $\lambda_i$.
It would be interesting to investigate whether the scatter of [Eu/Fe] can be interpreted as a distribution of the individual Lagrange parameter values that describe HEFO.

In the present work, it is not possible for us to assign values for the Lagrange parameters $\lambda_i$ to all the stars shown in the yellow box of Fig. \ref{EuFeBenjamin}, as this would require a considerable number of observed abundances of heavy elements; see \citet{2025FrASS..1233496B}.
We therefore carry out a rough estimate to determine the HEFO Lagrange parameters, using only the observed abundance $[{\rm Eu}/{\rm H}]$.
To derive the remaining two Lagrange parameters, we use the normalisation condition  $\sum_A X_A=1$.
We also assume the parameter $\lambda_T=4.4$ MeV.
This is an average value determined for low-metallicity halo stars considered in \citet{2025FrASS..1233496B}.
For a range of abundance ratios $[{\rm Eu}/{\rm H}]$, the values for the Lagrange parameters
are given in Table \ref{tab:1}.
Naturally, a more detailed investigation of the distribution of heavy elements in the stars considered here is required to determine the Lagrange parameters, in particular $\lambda_T$.

\begin{table}[htp]
\begin{center}
\begin{tabular}{|c|c|c|c|c|}
\hline
$[{\rm Eu}/{\rm Fe}]^{(0)}$& $[{\rm Fe}/{\rm H}]^{\rm HEFO}$ &$\lambda_T$ [MeV] & $\lambda_n$ [MeV] & $\lambda_p$ [MeV]  \\
\hline
 4&-4.702&4.4&940.99&844.20\\
3&-5.101&4.4&941.01&843.61\\
2&-5.499&4.4&941.03&843.02\\
1&-5.896&4.4&941.05&842.42\\
0&-6.292&4.4&941.07&841.83\\
\hline
\end{tabular}
\end{center}
\caption{Estimate of Lagrange parameters for low-metallicity stars, $[{\rm Fe}/{\rm H}]<-2$ for double fractions $[{\rm Eu}/{\rm Fe}]^{(0)}$, which is the value at $[{\rm Fe}/{\rm H}]=-4$, see Eq. (\ref{linFe}). A fixed value $\lambda_T=4.4$ MeV is assumed. $[{\rm Fe}/{\rm H}]^{\rm HEFO}$ is the double ratio of Fe and H at heavy-element freeze-out.}
\label{tab:1}
\end{table}%

\section{Discussion and Conclusions}
\label{conclusion}

We presented GCE calculations where the distribution of EMPS with very high $r$-process abundances can be reproduced by pre-enriched infall.
We show that our simulations are consistent with the observed data.
Our phenomenological approach anticipates the occurrence of $r$-process elements such as Eu already in old, metal-poor stars which represent the early Universe. 
This determines the initial conditions of the GCE calculations.
A corresponding distribution of the initial distribution of abundances [Eu/Fe] has been proposed.

The initial distribution of abundances [Eu/Fe] can be interpreted as local inhomogeneities in the early Universe which are smeared out during the temporary evolution by mixing. 
The resulting distribution of the Lagrange parameters can be considered as a signature for the assumption of
\citet{Wehmeyer:2015sra}  that local inhomogeneities can explain the observed large spread in the Eu abundances at low metallicities. 
Recent observations by the JWST \citep{2023ApJ...947L..24M} confirm the existence of large dense objects in the early Universe. \newline
To offer an interpretation of the results, and to suggest a possible site for the production of the needed pre-enriched infall, we suggest the HEFO scenario, based on the freeze-out concept which describes expanding hot and dense nuclear matter.
The heavy-element distribution freezes out very early as soon as the neutron-rich environment disappears.
The subsequent evolution of the heavy-element distribution is described by nuclear reaction networks.
The conditions at HEFO are described by Lagrange parameters, and we provided parameter values which are able to describe the observed unexpected high [Eu/Fe] rations for stars with very low [Fe/H].

The question of the origin of heavy elements remains unresolved at present
\citep{2021MNRAS.505.5862W,2024ApJ...971..158R,Bandyopadhyay:2025bzs,Arcones:2012wj,2025A&A...693A.294A,2023ApJ...958...45K,Cote:2016vla,2020ApJ...891L..13K}.
We propose an expansion of hot, dense nuclear matter in which a heavy-element freeze-out (HEFO) occurs as soon as the hot, neutron-rich environment ceases to exist.
The high $[{\rm Eu}/{\rm Fe}]$ ratio observed in some low-metallicity stars can be described phenomenologically using the derived HEFO Lagrange parameters $\lambda_T,\lambda_n,\lambda_p$.
Various sites come into consideration where the corresponding values for temperature, density and proton fraction are present, for example large, massive objects, mergers or the environment of primordial black holes.
The phenomenological freeze-out scenario could help to answer the question of why old, metal-poor stars already exhibit a considerable inventory of $r$-process elements such as Eu.
Natural expansions of this work would be, whether other heavy elements (such as barium as a tracer for the slow neutron capture process) would also be congruent with observations if HEFO is considered as a source of pre-enriched infall. This would enable us to compare their contribution ratios, e.g., [Eu/Ba] versus [Eu/H].

\begin{acknowledgments}
The work of BW was supported by the National Science Center (Narodowe Centrum Nauki, NCN, Poland) under grant No. 2022/47/D/ST9/03092. 
GR acknowledges a honorary stipend from the Foundation for Polish Science within the Alexander von Humboldt program under grant No. DPN/JJL/402-4773/2022.
This work was supported by the National Science Foundation (NSF, USA) under Grant no. OISE-1927130 (IReNA), the U.S. Department of Energy, Office of Nuclear Physics, under Award no. DE-SC0023128, as well as the COST action ``ChETEC-INFRA'' (G. A. no. 101008324).
The work of F.K.R. was supported by the Klaus Tschira Foundation and by the Deutsche Forschungsgemeinschaft (DFG, German Research Foundation) -- RO 3676/7-1, project number
537700965.  F.K.R.\ acknowledges funding by the European Union (ERC, ExCEED, project number 101096243). Views and opinions expressed are however those of the authors only and do not necessarily reflect those of the European Union or the European Research Council Executive Agency. Neither the European Union nor the granting authority can be held responsible for them.
Some computations outlined in this paper were performed at the Wroclaw Centre for Scientific Computing and Networking (WCSS).
\end{acknowledgments}

\section*{Data Availability Statement}
The observational datasets were extracted from the SAGA database (sagadatabase.jp, \citealt{Suda08,Suda11,Yamada13}). Observations were mainly taken from  \cite{Francois07,Simmerer04,Barklem05,Roederer10,Shetrone01,Shetrone03,Geisler05,Cohen09,Letarte10,Starkenburg13,McWilliam13}, under exclusion of carbon enhanced metal poor (``CEMP'') stars ([C/Fe]$\geq 1$ and [Fe/H]$\leq -1$), due to the high likelyhood of cross contamination in such stars. 
All data in this paper are available from the authors upon reasonable request.


\end{document}